\documentstyle[12pt,aasms4]{article}
\begin{document}
\title{DYNAMIC SCREENING AND THERMONUCLEAR REACTION RATES}
\author{Andrei V. Gruzinov}
\affil{Institute for Advanced Study, School of Natural Sciences,
Princeton, NJ 08540}

\begin{abstract}
We show that there are no dynamic screening corrections to the Salpeter's enhancement factor in the weak-screening limit.
\end{abstract}
\eject

\section{Introduction}
Continuing observations of solar neutrinos and helioseismological analyses of the solar interior (Bahcall 1989 and references therein) are becoming more accurate and warrant accurate theoretical expressions for thermonuclear reaction rates in astrophysical plasmas. As shown by Salpeter (1954), decreased electrostatic repulsion between reacting ions caused by the Debye-H\"uckel screening leads to a noticeable increase in reaction rates. The reaction rate enhancement factor is
\begin{equation}
w=\exp ({Z_1Z_2e^2\over TR_D}),
\end{equation}
where $R_D$ is the Debye radius. The above expression is valid in the weak-screening limit which is defined by $TR_D\gg Z_1Z_2e^2$. 

Carraro, Sch\"afer \& Koonin (1988) noted that if the Gamow energy of the reaction is high enough, reacting nuclei may have velocities much higher than thermal ion velocities. The plasma ions response will be suppressed, because at high enough velocities the ion part of the plasma dielectric constant is nearly zero (``dynamic screening''). The authors suggested that in such cases, only electron part of the screening should be included into (1), and Eq. (1) should read
\begin{equation}
w=\exp ({Z_1Z_2e^2\over \sqrt2 TR_D}),
\end{equation}
At even higher Gamow energies, the electron part of the dielectric constant also goes to zero, and, according to this argument, the screening enhancement effect disappears. 

We will show that although dynamic screening can be important for certain laboratory experiments, the nearly precise thermodynamic equilibrium of stellar plasmas guarantees that Salpeter's expression, Eq. (1), is valid independently of the Gamow energy (\S 3). Before that, in \S 2, we summarize the formal part of the argument of Carraro et al. (1988).

\section{What is Dynamic Screening?}

Consider a test charge $Z_1e$ moving through a plasma with velocity
$v$. The plasma response to the test particle is described by the
dielectric constant $\epsilon$ which is a function of both the wavenumber, $k$, and the frequency, $\omega$. For the purpose of reaction rate calculation, we need to know the electrostatic potential, $\phi _0$, created by the plasma at the test particle location. The enhancement factor is then (Salpeter, 1954)
\begin{equation}
w=\exp (-Z_2e\phi _0/T),
\end{equation}
for a $Z_1-Z_2$ reaction.
Now, knowing $\epsilon$ one calculates $\phi _0$ as
\begin{equation}
\phi _0=4\pi eZ_1\int {d^3k\over (2\pi)^3}( {1\over \epsilon(kv,k)}-1)k^{-2}.
\end{equation}
For $v=0$, we have $\epsilon =1+(kR_D)^{-2}$, and (4) gives  $\phi _0=-Z_1e/R_D$.
When inserted into Eq. (3), this result gives the Salpeter enhancement factor, Eq. (1). 

If reacting ions are fast moving, with velocities $v$ in the
interval $v_{Ti}<v<v_{Te}$, the dielectric constant in Eq. (4) is close to the static dielectric constant of the electron part only, that is $\epsilon \approx 1+(kR_D)^{-2}/2$. With this dielectric constant, Eq. (4) gives a $\sqrt2$ smaller potential $\phi _0$, which translates into a smaller enhancement factor, Eq. (2).

The above analysis seems well based, but is in contradiction with the following simple argument. The Gibbs probability distribution, $\rho$, for the plasma is
\begin{equation}
\rho \sim \exp (-\beta \sum {m_iv_i^2\over 2}-\beta \sum {e_ie_j\over r_{ij}}),
\end{equation}
where $\beta$ is the inverse temperature. This distribution is factorable,
\begin{equation}
\rho \sim \exp (-\beta \sum {m_iv_i^2\over 2})\times \exp (-\beta \sum {e_ie_j\over r_{ij}}).
\end{equation}
That is, distributions in velocity and configuration spaces are decoupled: fast
moving charges are just as screened as others. Salpeter's result, Eq. (1), can be derived starting from the probability distribution for spatial configurations, and thus does not depend on the Gamow energy.

\section{A Hypothetical Reaction}
The strongest disagreement between Salpeter's expression and the dynamic screening prediction would occur in an extreme case of very high Gamow energy. Then interacting nuclei are moving at very high velocities, and the plasma does not respond at all. We will show that even in this case Salpeter's expression is correct.

The plasma does not respond to fast moving nuclei $Z_1$ and $Z_2$, but the nuclei move in a thermal (fluctuating) electrostatic field, $\phi (r)$. We will show that reaction rates in a random potential are increased, and for thermal fluctuations this effect is correctly accounted for by the Salpeter expression. 

In thermal equilibrium, the local densities of the nuclei are
\begin{equation}
n_{1,2}(r)=C_{1,2}\exp (-\beta Z_{1,2}e\phi (r)).
\end{equation}
The local reaction rate is $Kn_1(r)n_2(r)$, where $K$ is a multiplicative factor that depends only on temperature. The average reaction rate is, for a gaussian random field $\phi$,
\begin{equation}
R~=~K<n_1(r)n_2(r)>~=~KC_1C_2\exp ({1\over 2}\beta ^2e^2(Z_1+Z_2)^2<\phi ^2>),
\end{equation}
where we have taken $<\phi >=0$ with no loss of generality.
Now, the average densities of the reacting nuclei are related to the normalization constants $C_{1,2}$ by
\begin{equation}
<n_{1,2}>~=~C_{1,2}\exp ({1\over 2}\beta ^2e^2Z_{1,2}^2<\phi ^2>),
\end{equation}
and, in terms of mean densities, Eq. (8) reads
\begin{equation}
R~=~K<n_1><n_2>\exp (\beta ^2e^2Z_1Z_2<\phi ^2>).
\end{equation}
This corresponds to an enhancement factor
\begin{equation}
w=\exp (\beta ^2e^2Z_1Z_2<\phi ^2>).
\end{equation}

Equation (11) is valid for an arbitrary gaussian random potential. To make a connection with Salpeter's formula, we have to calculate the size of the thermal electrostatic fluctuations, $<\phi ^2>$. The fluctuating electric field, $E$, is given by the equipartition law, 
\begin{equation}
{<E^2>_k\over (8\pi )}=(T/2)(1-{1\over \epsilon(0,k)}),
\end{equation}
where the notation $<E^2>_k$ is defined by
\begin{equation}
<E^2>=\int {d^3k\over (2\pi )^3}<E^2>_k.
\end{equation}
The potential fluctuation is $<\phi ^2>_k=<E^2>_k/k^2$, and Eq. (12) gives
\begin{equation}
<\phi ^2>_k=(4\pi T)k^{-2}(1+k^2R_D^2)^{-2}.
\end{equation}
Now, 
\begin{equation}
<\phi ^2>=\int {d^3k\over (2\pi )^3}<\phi ^2>_k={T\over R_D},
\end{equation}
and Eq. (11) gives 
\begin{equation}
w=\exp ({Z_1Z_2e^2\over TR_D}).
\end{equation}
This is exactly Salpeter's result. It was obtained without explicitly introducing the concept of screening.

\acknowledgements

I thank John Bahcall and Marshall Rosenbluth for helpful 
discussions and suggestions. This work was supported by NSF PHY-9513835.
\vfil\eject

\end{document}